\documentclass[english]{article}
\usepackage[T1]{fontenc}
\usepackage[latin9]{inputenc}
\usepackage{geometry}
\usepackage{float}
\usepackage{bm}
\usepackage{amstext}
\usepackage{amssymb}
\usepackage{graphicx}

\makeatletter
\usepackage{babel}

\makeatother

\usepackage{babel}
\begin{document}

\title{On the role of dissipation in structure formation for dilute relativistic
gases: the static background case}

\author{A. R. Méndez$^{1}$, A. L. García-Perciante$^{2}$ and A. Sandoval-Villalbazo$^{3}$}

\date{{\small{}$^{1,2}$Depto. de Matemáticas Aplicadas y Sistemas, Universidad
Autónoma Metropolitana-Cuajimalpa, Prol. Vasco de Quiroga 4871, México
D.F 05348, México.}\\
{\small{}$^{3}$Depto. de Física y Matemáticas, Universidad Iberoamericana,
Prolongación Paseo de la Reforma 880, México D. F. 01219, México.}}
\maketitle
\begin{abstract}
A correction to the Jeans stability criterion due to heat conduction
is established for the case of high temperature gases. This effect
is only relevant for relativistic fluids and includes an additional
term due to a density gradient driven heat flux. The result is obtained
by thoroughly analyzing the exponentially growing modes present in
the dynamics of density fluctuations in the linearized relativistic
Navier-Stokes regime. The corrections to the corresponding Jeans mass
and wavenumber are explicitly obtained and are compared to the non-relativistic
and non-dissipative values using the transport coefficients obtained
in the BGK approximation. 
\end{abstract}

\section*{Introduction }

The establishment of conditions for the onset of gravitational instabilities
in a dilute gas is one of the most important problems in astrophysics.
Transport theory can be directly applied to tackle this problem using
a simplified model in which the field of an isothermal self-gravitating
gas surpasses its corresponding hydrostatic pressure in the absence
of dissipation \cite{Jeans}. This basic model leads to the well-known
critical parameters for gravitational structure formation namely,
\textit{Jeans wave number} and \textit{Jeans mass}. These parameters
are the standard benchmarks to understand the characteristic lengths
and formation times corresponding to fundamental astrophysical objects
such as stars and galaxies. It is natural to analyze the problem of
the growth of density fluctuations due to gravity including several
effects such as cosmological expansion. Pioneering work related to
entropy production in the realm of cosmology was performed over 40
years ago by Weinberg \cite{Weinberg} and was later included in classical
textbooks \cite{Kolb-Turner}. The introduction of dissipation was
first analyzed using numerical stability techniques by Corona-Galindo
and Dehnen \cite{Corona-Galindo}; later on, useful algebraic representations
for the dispersion relation governing the evolution of density fluctuations
were developed, identifying a non-oscillating damped mode due to viscous
processes \cite{Carlevaro-Montani2009}.

The introduction of heat conduction in the analysis of Jeans instability
in the case of a high temperature (relativistic) dilute fluid, was
carefully studied in Ref. \cite{Mondragon-Sandoval}, identifying
serious problems with Eckart's constitutive equation (which proposes
a coupling of the heat flux with the hydrodynamic acceleration of
the fluid). Along with the development of relativistic kinetic theory,
a relativistic constitutive equation up to first order in the gradients
was successfully established \cite{Sandoval-Garcia-Garcia2009} in
order to eliminate the generic instabilities first identified by Hiscock
and Lindblom \cite{Hiscock-Lindblom} that were responsible of the
pathological features mentioned in Ref \cite{Mondragon-Sandoval}.
Two damped non-oscillatory modes due to viscosity can also be found
in the relativistic case, but the effects of heat conduction to the
actual modified values of the Jeans wave number have not been addressed
before in the case of first order in the gradients constitutive equations.
It is the purpose of this work to fill this gap, showing that the
first dissipative effects affecting real growing density modes in
a simple relativistic fluid \textit{are those associated to heat conduction},
instead of viscosity. These rather surprising result is obtained as
follows: In section 2 the relativistic transport equations for self-gravitating
dilute fluids in the Navier-Stokes regime are reviewed. The linearized
version of the transport system is obtained in section 3 leading to
an algebraic dispersion relation which yields the relativistic Jeans
wave-number in the non-dissipative limit. The conditions for gravitational
collapse including heat and viscosity are thoroughly discussed in
section 4, so that a simple dimensionless parameter involving the
heat conductivity coefficient is established in order to evaluate
the effect of heat conduction to the ordinary Jeans wave number. A
discussion of this result and final remarks are included in the final
section of this work.

\section*{The relativistic transport equations}

As mentioned above, the system here addressed is a self gravitating
dilute neutral gas in a relativistic, non-equilibrium situation. The
relativistic nature of such system arises from the fact that the temperature
is high enough such that the ratio of thermal to individual molecules'
rest energy is non-negligible, leading to relevant relativistic corrections
in the evolution of state variables. In this sense, full relativistic
effects are considered for the system, however since the density of
the gas is assumed to be low, the gravitational potential is only
considered within a linear approximation. Thus, in the case of a static
background, the corresponding metric reads 
\begin{equation}
ds^{2}=dr^{2}+r^{2}\left(d\theta^{2}+\sin\theta d\varphi^{2}\right)-\left(1-\frac{2\phi\left(r\right)}{c^{2}}\right)c^{2}dt^{2}.\label{eq:metric}
\end{equation}
where $\phi$ is the gravitational potential and $c$ the speed of
light.

In order to address the response of the gas to a density fluctuation,
the relativistic transport equations need to be analyzed under such
conditions. Based on relativistic kinetic theory, it has been shown
that such equations, considering the number density $n$, the internal
energy $\varepsilon$ and the hydrodynamic four-velocity $u^{\nu}$
as state variables, are given by two conservation statements in the
corresponding four-dimensional spacetime, namely \cite{KremerLibro,microscopic}
\begin{equation}
N_{;\mu}^{\text{\ensuremath{\mu}}}=0\label{eq:particle}
\end{equation}
and 
\begin{equation}
T_{;\nu}^{\mu\nu}=0,\label{eq:energy}
\end{equation}
which represent particle and energy-momentum balances, respectively.
Here $N^{\mu}=nu^{\mu}$ is the particle four-flow and 
\[
T_{\nu}^{\mu}=\frac{n\varepsilon}{c^{2}}u^{\mu}u_{\nu}+ph_{\nu}^{\mu}+\pi_{\nu}^{\mu}+\frac{1}{c^{2}}q^{\mu}u_{\nu}+\frac{1}{c^{2}}u^{\mu}q_{\nu},
\]
is the stress-energy tensor, where $p$ is the hydrostatic pressure,
$q_{\nu}$ is the heat flux, $\pi_{\nu}^{\text{\ensuremath{\mu}}}$
is the Navier tensor and $h^{\mu\text{\ensuremath{\nu}}}$ is the
usual spatial projector. By substituting $N^{\mu}$ and $T_{\nu}^{\mu}$
in Eqs. (\ref{eq:particle}) and (\ref{eq:energy}), the set of transport
equations can be shown to be explicitly given by

\begin{equation}
\dot{n}+n\theta=0,\label{eq:t1}
\end{equation}
\begin{equation}
\left(\frac{n\varepsilon}{c^{2}}+\frac{p}{c^{2}}\right)\dot{u}_{\nu}+\left(\frac{n\dot{\varepsilon}}{c^{2}}+\frac{p}{c^{2}}\theta\right)u_{\nu}+p_{,\mu}h_{\nu}^{\mu}+\pi_{\nu;\mu}^{\mu}+\frac{1}{c^{2}}\left(q_{;\mu}^{\mu}u_{\nu}+q^{\mu}u_{\nu;\mu}+\theta q_{\nu}+u^{\mu}q_{\nu;\mu}\right)=0,\label{eq:t2}
\end{equation}
\begin{equation}
nC_{n}\dot{T}+p\theta+u_{;\mu}^{\nu}\pi_{\nu}^{\mu}+q_{;\mu}^{\mu}+\frac{1}{c^{2}}\dot{u}^{\nu}q_{\nu}=0,\label{eq:t3}
\end{equation}
where a semicolon denotes a covariant derivative and a dot a proper
time derivative $\dot{A}_{\nu}=u^{\mu}A_{\nu;\mu}$. Also, $\theta=u_{;\nu}^{\nu}$
and $C_{n}$ is the heat capacity at constant particle density. Notice
that here the gravitational field acts on the fluid through space-time
curvature and thus, the gravitational potential is present in Eq.
(\ref{eq:t2}) through the covariant derivatives 
\begin{equation}
A_{\nu;\mu}=A_{\nu,\mu}-\Gamma_{\mu\nu}^{\lambda}A_{\lambda},\label{eq:derivative}
\end{equation}
where $\Gamma_{\mu\nu}^{\lambda}$ are the usual Christoffel symbols.

As is well known, the set of transport equations given by Eqs. (\ref{eq:t1})-(\ref{eq:t3})
requires closure relations for the dissipative fluxes $\mathcal{Q}^{\nu}$and
$\pi^{\mu\nu}$ and an additional equation for the gravitational potential.
The constitutive equations that relate the dissipative fluxes with
the state variables' gradients are given by the expressions obtained
from relativistic kinetic theory, namely \cite{Sandoval-Garcia-Garcia2009,bulk2015}
\begin{equation}
\pi_{\nu}^{\mu}=-2\eta h_{\alpha}^{\mu}h_{\nu}^{\beta}\tau_{\beta}^{\alpha}-\zeta\theta\delta_{\nu}^{\mu}\label{eq:viscous}
\end{equation}
and 
\begin{equation}
\mathcal{Q}^{\nu}=-h_{\mu}^{\nu}\left(L_{TT}\frac{T^{,\mu}}{T}+L_{nT}\frac{n^{,\mu}}{n}\right).\label{eq:heat}
\end{equation}
The explicit expressions for the transport coefficients $\eta,$ $\zeta,$
$L_{TT},$ $L_{nT}$ in a relaxation time approximation can be found
in Refs. \cite{Sandoval-Garcia-Garcia2009,KremerLibro,bulk2015} (also
see the Appendix). Meanwhile, for the gravitational field, a Poisson
equation 
\begin{equation}
\nabla^{2}\phi=-4\pi G\rho\label{eq:poisson}
\end{equation}
is considered. Introducing Eqs. (\ref{eq:viscous})-(\ref{eq:poisson})
in Eqs. (\ref{eq:t1})-(\ref{eq:t3}) closes the set for the variables
$n$, $u^{\nu}$ and $T$. In the next section, the effects of the
relativistic terms contained such set in the conditions for a gravitational
collapse will be explored within the linear approximation.

\section*{Density fluctuations}

In dilute self-gravitating gases, a gravitational collapse may be
initiated provided the conditions for the exponential growth of density
fluctuations are met for the particular wavelength of such perturbations.
The standard approach to this problem consists in assuming small deviations
from local equilibrium values for the state variables and analyzing
their dynamics in a linear approximation. Thus, we assume a given
state variable $X$ can be written as 
\begin{equation}
X\left(\bm{r},t\right)=X_{0}+\delta X\left(\bm{r},t\right),\label{eq:statevariables}
\end{equation}
where $X_{0}$ is its equilibrium value and $\delta X$ a fluctuation
given by 
\[
\delta X\left({\bf r},t\right)=\hat{X}e^{-i\bm{{\bf q}}\cdot{\bf \bm{r}}+st},
\]
with ${\bf k}$ being the wave number vector, $k$ its magnitude,
$s$ the (real) angular frequency and $\hat{X}$ a constant amplitude.
Introducing such assumption in the set of transport equations and
considering up to linear terms in fluctuations, the following set
is obtained 
\begin{equation}
\frac{\partial\left(\delta n\right)}{\partial t}+n_{0}\delta\theta=0,\label{eq:t1-2}
\end{equation}
\begin{equation}
\tilde{\rho}_{0}\frac{\partial\left(\delta\theta\right)}{\partial t}+p_{0}\left(\frac{\nabla^{2}\left(\delta T\right)}{T_{0}}+\frac{\nabla^{2}\left(\delta n\right)}{n_{0}}\right)-A\nabla^{2}\left(\delta\theta\right)-\frac{1}{c^{2}}\left(L_{TT}\frac{\nabla^{2}\left(\delta\dot{T}\right)}{T_{0}}+L_{nT}\frac{\nabla^{2}\left(\delta\dot{n}\right)}{n_{0}}\right)=-4\pi Gm\tilde{\rho}_{0}\delta n,\label{eq:t2-2}
\end{equation}
\begin{equation}
C_{n}n_{0}\frac{\partial\left(\delta T\right)}{\partial t}+p_{0}\delta\theta-\frac{L_{TT}}{T_{0}}\nabla^{2}\left(\delta T\right)-\frac{L_{nT}}{n_{0}}\nabla^{2}\left(\delta n\right)=0,\label{eq:t3-2}
\end{equation}
where $p_{0}=n_{0}kT_{0}$ . Also 
\[
\tilde{\rho}_{0}=\frac{n_{0}\varepsilon_{0}+p_{0}}{c^{2}}=n_{0}m\mathcal{G}\left(\frac{1}{z}\right),
\]
where $\mathcal{G}\left(1/z\right)=\mathcal{K}_{3}\left(1/z\right)/\mathcal{K}_{2}\left(1/z\right)$
with $\mathcal{K}_{n}$ being the $n$-th modified Bessel function
of the second kind and $A=\eta+4\xi/3$. Notice that Eq. (\ref{eq:t2-2})
corresponds to the divergence of the momentum balance equation. Since
both the continuity and energy balance equations are coupled with
the velocity only through its divergence $\delta\theta$, and all
the source terms in Eq. (\ref{eq:t2}) are given by gradients, the
evolution equation for the curl of the velocity field is not coupled
to the system and a set of three scalar equations is obtained. This
transverse mode decays exponentially with time \cite{Sandoval-Garcia-Garcia2009}.

The set of linearized equations (\ref{eq:t1-2})-(\ref{eq:t3-2}),
following the standard procedure, is analyzed in Laplace-Fourier space
where they can be written in matrix form, where the inverse susceptibility
matrix $M$, is given by 
\[
M=\left(\begin{array}{ccc}
s & n_{0} & 0\\
-kT_{0}q^{2}+L_{nT}sq^{2}/n_{0}c^{2}+4\pi Gm\tilde{\rho}_{0} & Aq^{2}+\tilde{\rho}_{0}s & -n_{0}kq^{2}+L_{TT}sq^{2}/T_{0}c^{2}\\
L_{nT}q^{2}/n_{0} & n_{0}kT_{0} & C_{n}n_{0}s+L_{TT}q^{2}/T_{0}
\end{array}\right).
\]
The corresponding dispersion relation can be written as a third order
polynomial as follows 
\begin{eqnarray}
s^{3}+\left(\frac{A}{\tilde{\rho}_{0}}+\frac{L_{TT}}{n_{0}TC_{n}}-D_{TR}\right)q^{2}s^{2}+\left(\frac{AL_{TT}k}{\tilde{\rho}_{0}p_{0}C_{n}}q^{4}+\frac{p_{0}}{\tilde{\rho}_{0}}q^{2}\left(1+\frac{k}{C_{n}}\right)-4\pi Gmn_{0}\right)s\label{eq:disp}\\
+\frac{k}{C_{n}\tilde{\rho}_{0}}\left(L_{TT}-L_{nT}\right)q^{4}-\frac{4\pi GmL_{TT}}{T_{0}C_{n}}q^{2}=0,\nonumber 
\end{eqnarray}
where 
\begin{equation}
D_{TR}=\frac{1}{\tilde{\rho}_{0}c^{2}}\left(L_{nT}+\frac{k}{C_{n}}L_{TT}\right),\label{eq:dtr}
\end{equation}
is a generalized thermal diffusivity. In order to obtain approximate
solutions for $s$ it is assumed, as in previous works, that equation
(\ref{eq:disp}) has three different roots and can thus be factored
as 
\begin{equation}
\left(s-\gamma q^{2}\right)\left(s^{2}+\mu s+\nu\right)=0.\label{eq:disp-fact}
\end{equation}
To order $q^{2}$ one can show that the real root is not affected
by the gravitational field and is given by $\gamma=-L_{TT}/n_{0}T_{0}C_{n}$.
This corresponds to a decaying mode and is associated with Rayleigh's
peak when a scattering spectrum is present. On the other hand, the
coefficients $\mu$ and $\nu$ feature both relativistic and gravitational
corrections: 
\begin{equation}
\mu=\frac{q^{2}L_{TT}}{\tilde{\rho}_{0}}\left(\frac{A}{L_{TT}}-\frac{L_{nT}}{c^{2}L_{TT}}-\frac{k}{c^{2}C_{n}}\right)\label{eq:mu}
\end{equation}
and 
\begin{equation}
\nu=-4\pi Gmn_{0}+\frac{p_{0}}{\tilde{\rho}_{0}}\left(1+\frac{k}{C_{n}}\right)q^{2}+\frac{k}{C_{n}}\frac{L_{TT}D_{TR}}{p_{0}\tilde{\rho}_{0}c^{2}}q^{4}.\label{eq:nu}
\end{equation}
The values of $\mu$ and $\nu$ determine the dynamics of the fluctuations.
In the non-dissipative case $\mu=0$ and the fluctuations either oscilate
in the case of purely imaginary values for $s$ or feature a combination
of growing and decaying modes, depending on the value of $q$. In
that case, it is straightforward to obtain that 
\[
\tilde{q}_{J}^{2}=\frac{4\pi Gn_{0}m}{C_{s}^{2}},
\]
where $C_{s}^{2}=zc^{2}\left(1+k/C_{n}\right)/\mathcal{G}\left(1/z\right)$
is the relativistic speed of sound, marks the boundary between purely
oscillatory ($q^{2}<\tilde{q}_{J}^{2}$) and exponential ($q^{2}>\tilde{q}_{J}^{2}$)
behavior. Notice that $\tilde{q}_{J}$ corresponds to the relativistic
value of the Jean's wave number corrected by relativistic the value
of $C_{s}$.

Returning to Eq. (\ref{eq:disp-fact}), since $\mu\neq0$, the purely
oscillatory behavior is not present. This occurs also in the dissipative
non-relativistic case \cite{Carlevaro-Montani2009}. If the quadratic
expression in (\ref{eq:disp-fact}) has two real roots, the density
fluctuations either grow of decay with time. On the other hand, if
the roots have a non-vanishing imaginary part, and since $\mu>0$,
the dynamics of the fluctuations correspond to damped oscillations
which can be identified with the Brillouin peaks in the corresponding
spectrum \cite{GPercianteGColinSandoval-RBSpectrum}.

The case of interest in this work is the one leading to an exponential
growth of $\delta n$, which could iniciate to a gravitational collapse
of the cloud. The condition on the wave number, depending on the dissipation
parameters, for this phenomenon to be feasible in the system will
be analyzed in the next section. It is very important to point out
at this stage that the quartic order term in equation (\ref{eq:nu}),
which is usually neglected when comparing $\mu^{2}$ and $\nu$ and
is here retained, arises from the purely relativistic term in Eq.
(\ref{eq:t2}), which indicates the presence of thermal dissipation
of momentum for high temperature gases.

\section*{Gravitational collapse}

In this section, a detailed analysis of the factor $\left(s^{2}+\mu s+\nu\right)$
in the dispersion relation is carried out in order to obtain the criterion
for a gravitational instability to occur. Clearly, the limiting value
which separates exponential and oscillatory regimes for density fluctuations
is determined by the sign of $\bar{s}=1-\frac{4\nu}{\mu^{2}}$ since
the corresponding roots are given by 
\begin{equation}
s=-\frac{\mu}{2}\left(1\pm\sqrt{\bar{s}}\right).\label{eq:s}
\end{equation}
Imaginary solutions are obtained for $\bar{s}<0$ which, since $\mu>0$
(see Fig. 1), leads to damped oscillations. On the other hand, for
$\bar{s}>0,$ one obtains real roots and thus an exponential time
dependence for $\delta n$, which corresponds to either growing or
decaying modes. It is important to notice that the criterion for exponential
growth is not completely given by $\bar{s}>0$, but also by the sign
of $s$ since only real and \emph{positive} values of $s$ may lead
to a gravitational collapse. Using once again the fact that $\mu>0$,
it is clear that such requirement is met only if $\bar{s}>1$ (considering
the ``$-$'' sign in Eq. (\ref{eq:s})), or equivalently $\nu<0$.

\begin{figure}[H]
\begin{centering}
\includegraphics[scale=0.6]{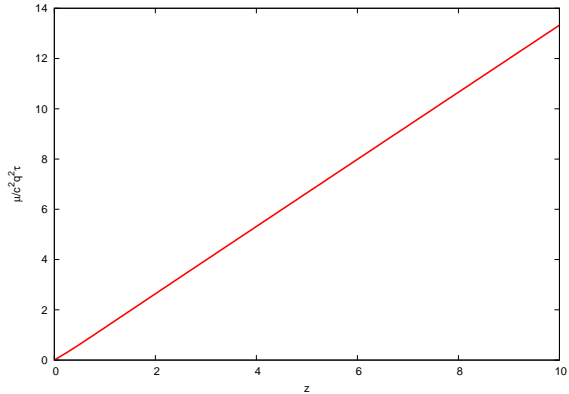} 
\par\end{centering}

\centering{}\protect\protect\caption{\label{fig:1} The quantity $\mu$ in Eq. (\ref{eq:mu}) as a function
of $z$, here we have used the explicit expressions form $L_{nT}\left(z\right),L_{TT}\left(z\right),\eta\left(z\right)$
and $\zeta\left(z\right)$ \cite{Sandoval-Garcia-Garcia2009,KremerLibro,bulk2015}.}
\end{figure}

Thus, considering $\nu=0$ as the equation determining the Jeans criterion,
the corresponding critical wave number can be written as 
\begin{equation}
q_{J}^{2}=\frac{C_{s}^{2}n_{0}T_{0}C_{n}}{2L_{TT}D_{TR}}\left(\sqrt{1+\frac{16\pi GmL_{TT}}{T_{0}C_{n}}\frac{D_{TR}}{C_{s}^{4}}}-1\right).\label{eq:qj2}
\end{equation}
Imaginary solutions have been excluded in Eq. (\ref{eq:qj2}) since
only temporal instabilities are relevant for this problem. The expression
for $q_{J}$ above corresponds to the relativistic Jeans' wave number
in the case of dissipative fluids and constitutes the main contribution
of this work. Notice that, eventhough the viscosity plays a role in
separating imaginary and real solutions, the final criterion for growing
modes is independent of $A$. Indeed, since $\bar{s}>1$ is enforced,
the condition that $\bar{s}$ be positive remains weaker and is already
satisfied independently of the values of the viscosities. The fact
that viscous dissipation does not alter the value for $q_{J}$ was
already pointed out, for non-relativistic systems ($z\ll1$), in Ref.
\cite{Carlevaro-Montani2009}. Thus, dissipation affects the Jeans
criterion only for high temperature gases and in that case, it is
only thermal dissipation that plays a role.

\begin{figure}[H]
\includegraphics[scale=0.6]{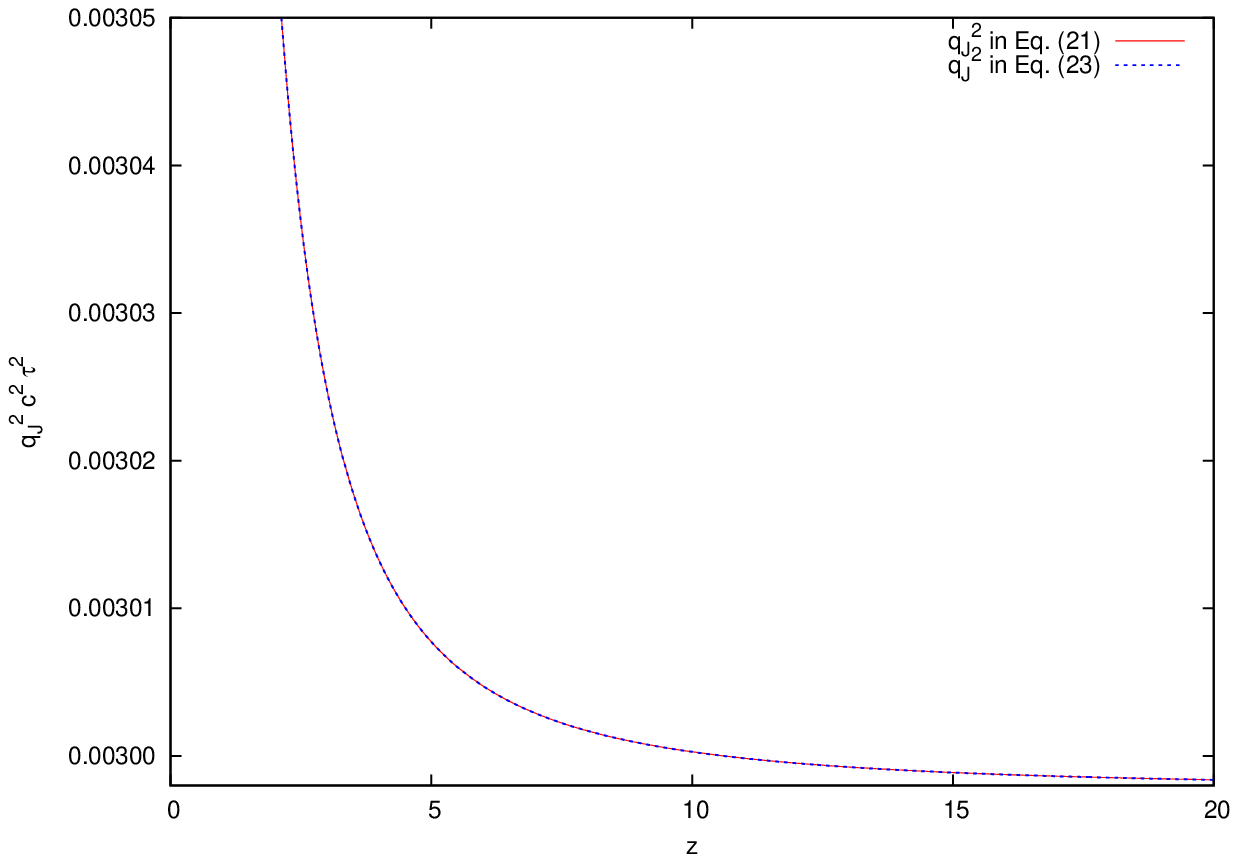}\includegraphics[scale=0.6]{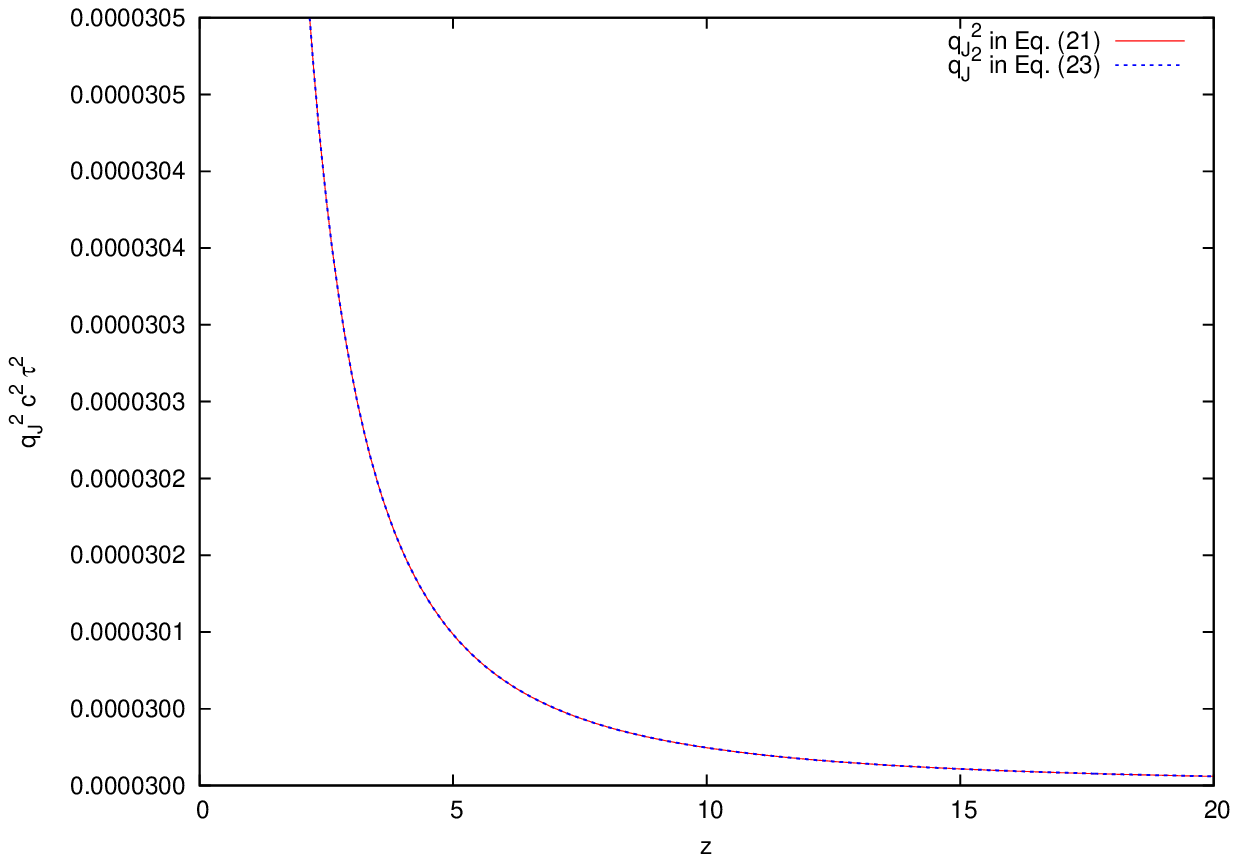}\protect\protect\caption{\label{fig:2}Jeans wave number in Eq. (\ref{eq:qj2}) and its approximation
in Eq. (\ref{eq:qc2}) for $R$=0.001 at the left and $R=0.00001$
at the right. }
\end{figure}

To get some insight on the nature of the relativistic corrections
arising from dissipative effects we consider 
\begin{equation}
\epsilon\equiv\frac{16\pi GmL_{TT}}{T_{0}C_{n}}\frac{D_{TR}}{C_{s}^{4}}\ll1\label{eq:assume}
\end{equation}
in Eq. (\ref{eq:qj2}), such that the Jeans' number can be written
as (see Figure \ref{fig:2}) 
\begin{equation}
q_{J}^{2}\approx\tilde{q}_{J}^{2}\left(1-\frac{\epsilon}{4}\right).\label{eq:qc2}
\end{equation}
Notice that, since in the relaxation approximation all transport coefficients
are proportional to the relaxation time $\tau$ \cite{KremerLibro,Sandoval-Garcia-Garcia2009,bulk2015},
the correction $\epsilon$ results in a dimensionless parameter that
depends on the ratio of the microscopic and gravitational characteristic
times. Defining $R=\left(\tau/\tau_{g}\right)^{2}$, with $\tau_{g}=1/\sqrt{4\pi Gmn}$,
being the gravitational time scale, Fig. (2) verifies the validity
of the approximation in Eq. (\ref{eq:qc2}) for two values of $R$.

From these results one can also determine the Jeans' mass 
\begin{equation}
M_{J}=\frac{4}{3}\pi\rho\left(\frac{\pi}{q_{J}}\right)^{3},\label{eq:mj}
\end{equation}
which under the assumption $\epsilon\ll1$ can be written as 
\begin{equation}
M_{J}\approx\tilde{M}_{J}\left(1+\frac{3}{8}\epsilon\right).\label{eq:mjc}
\end{equation}
Here $\tilde{M}_{J}=\frac{4}{3}\pi\rho\left(\pi/\tilde{q}_{J}\right)^{3}$
is the relativistic Jeans mass in the absence of dissipation.

The magnitude of the correction to the Jeans' mass and wavenumber,
given essentially by the parameter $\epsilon$, can be appreciated
in Fig. (2). The plot shows the dependence of $\epsilon$ with the
relativistic parameter $z=kT/mc^{2}$, considering the explicit expressions
for the transport coefficients $L{}_{n}$, $L_{T}$, $\eta$ and $\zeta$
obtained in \cite{Sandoval-Garcia-Garcia2009,KremerLibro,bulk2015}
(see the Appendix) and for two different values of $R$. Notice that
the correction vanishes in the non-relativistic limit and reaches
a constant value for large $z$. Indeed, $\epsilon$ can be written
as (see Appendix) 
\[
\epsilon\equiv\mathcal{F}\left(z\right)R
\]
where $\lim_{z\rightarrow0}\mathcal{F}\left(z\right)=0$ and $\lim_{z\rightarrow0}\mathcal{F}\left(z\right)=3$.

\begin{figure}[H]
\includegraphics[scale=0.6]{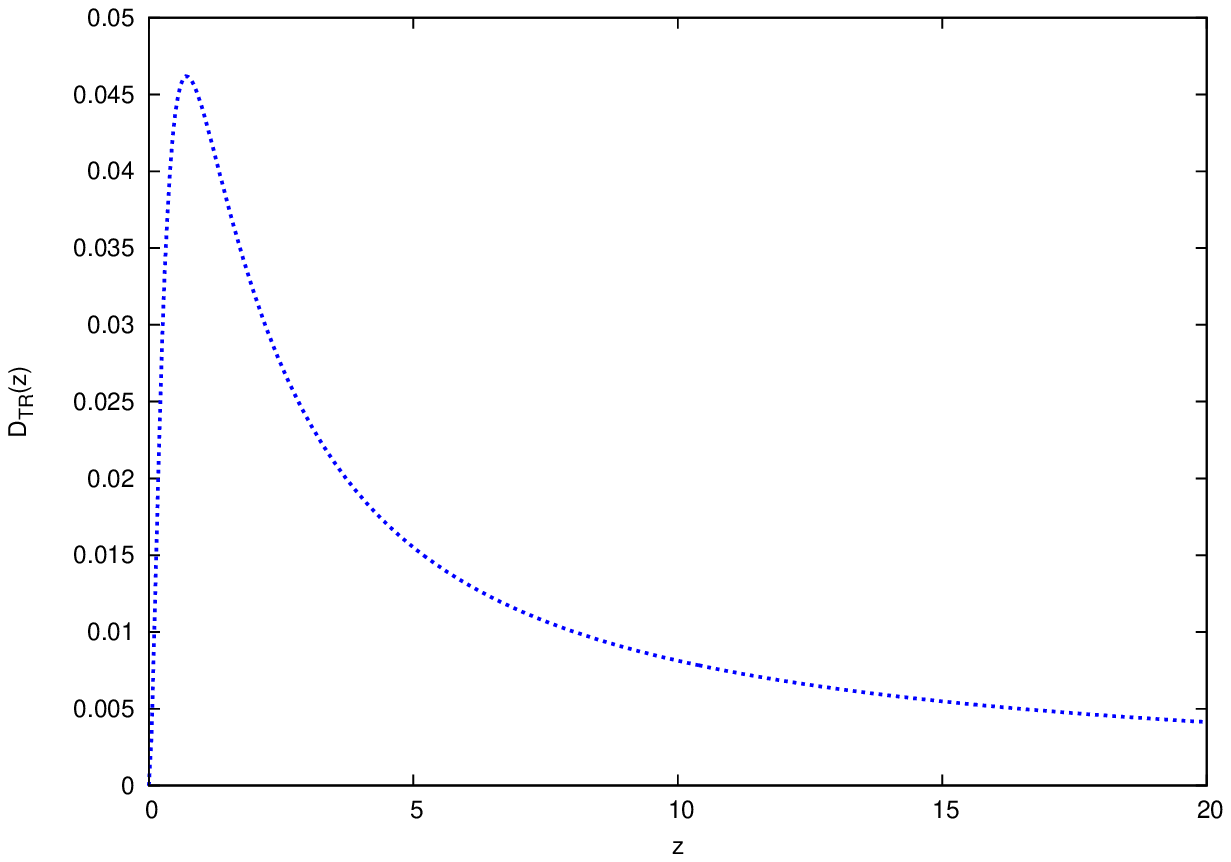}\includegraphics[scale=0.6]{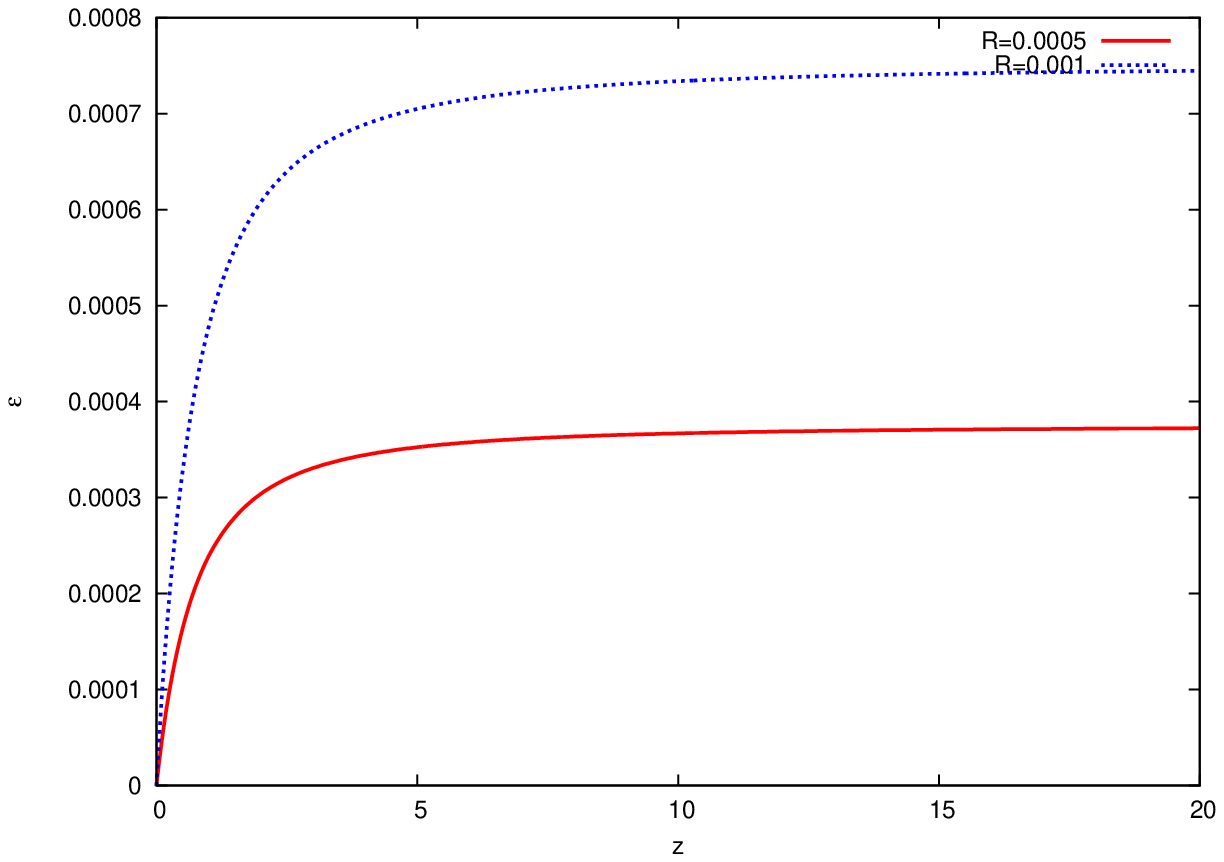}

\protect\protect\caption{\label{fig:3}The quantities $D_{TR}$ and $\epsilon$ in Eqs. (\ref{eq:dtr})
and (\ref{eq:assume}) respectively.}
\end{figure}

\section*{Discussion and concluding remarks}

The generalization to the Jeans' criterion for high temperature dissipative
systems was obtained as expressions for $q_{J}$ and $M_{J}$. These
parameters depend on the value of the transport coefficients as well
as the mass of the system. Eventhough they are closely related, one
can try to separate the physical origin of the new values on the purely
relativistic and the dissipative ones. The isolated effect of the
high temperature impacts the values of the inertia coefficient $\rho_{0}\rightarrow\tilde{\rho}_{0}$
which is actually rooted in the relativistic expression for the internal
energy. Also the heat capacity $C_{n}$ and the speed of sound $C_{s}$
are modified for $z\gtrsim1$.

On the other hand, the effects of dissipation arise from the relativistic
heat terms in the momentum balance equation. In particular, these
effects can be traced back to the last term on the left hand side
of Eq. (\ref{eq:t2}). It is worthwhile to mention that this same
term was found to be initially the source of the so-called generic
instabilities when a constitutive equation featuring an acceleration
term instead of the density gradient term in Eq. (\ref{eq:heat})
was considered \cite{Hiscock-Lindblom}. A detailed discussion of
this point can be found in Ref. \cite{Sandoval-Garcia-Garcia2009}.
One can thus argue that this particular term introduces significant
modifications to the dynamics of the relativistic fluid, more so than
the relativistic corrections to transport coefficients and other parameters.
Moreover, as can be seen by inspecting the set of linearized transport
equations (Eqs. (\ref{eq:t1-2})-(\ref{eq:t3-2})), thermal dissipation
due to density and temperature gradients enhance the effects of the
pressure gradient, thus increasing the mass required for the collapse.
More precisely, as can be seen in Fig. (3), the parameter $D_{TR}$,
which features a combination of both transport coefficients, is positive
for all values of $z$. It is important to emphasize at this point
that only in the relativistic regime, and due to the relativistic
heat terms in Eq. (\ref{eq:t2-2}), dissipation alters the Jeans'
criterion. For dissipative non-relativistic fluids ($z\ll1$), $q_{J}$
and $M_{J}$ have the same values as in the non-dissipative case.

\section*{Acknowledgements}

The authors acknowledge support from CONACyT through grant number
CB2011/167563.

\section*{Appendix }

In this section we show explicitly the transport coefficients appearing
in the closure relations (\ref{eq:viscous}) and (\ref{eq:heat}).
The quantities $\eta\left(z\right),\xi\left(z\right),L_{TT}\left(z\right)$
and $L_{nT}\left(z\right)$ correspond to the bulk viscosity, shear
viscosity and the conductivities associated to heat flux respectively,
as functions of the relativistic parameter $z$. These coefficients
where obtained through the usual methods in kinetic theory based on
a relaxation time approximation \cite{Sandoval-Garcia-Garcia2009,KremerLibro,bulk2015}.
For the viscosities we have 
\begin{equation}
\eta\left(z\right)=\frac{1}{3}nmc^{2}\tau\frac{\left[z\left(2-20z^{2}\right)\mathcal{G}\left(\frac{1}{z}\right)+13z^{2}\mathcal{G}^{2}\left(\frac{1}{z}\right)-2z\mathcal{G}^{3}\left(\frac{1}{z}\right)-3z^{2}\right]}{3z^{2}\mathcal{G}\left(\frac{1}{z}\right)\alpha\left(z\right)}\label{eq:eta}
\end{equation}
and 
\begin{equation}
\xi\left(z\right)=nmc^{2}\tau z\mathcal{G}\left(\frac{1}{z}\right),\label{eq:xi}
\end{equation}
where 
\begin{equation}
\alpha\left(z\right)=\frac{1}{z}\left[-\frac{1}{z}\mathcal{G}\left(\frac{1}{z}\right)+\frac{1}{z\mathcal{G}\left(\frac{1}{z}\right)}+5\right]-\frac{1}{\mathcal{G}\left(\frac{1}{z}\right)}.\label{eq:alfa}
\end{equation}
Considering that heat flux is written as in Eq. (\ref{eq:heat}) we
have

\begin{equation}
L_{TT}\left(z\right)=nmc^{4}z^{3}\tau\left[\frac{1}{z}\mathcal{G}\left(\frac{1}{z}\right)-1\right]\left[\alpha\left(z\right)+\frac{1}{\mathcal{G}\left(\frac{1}{z}\right)}\right]\label{eq:ltt}
\end{equation}
and 
\begin{equation}
L_{nT}\left(z\right)=-nmc^{4}z^{3}\tau\left[\alpha\left(z\right)+\frac{1}{\mathcal{G}\left(\frac{1}{z}\right)}\right],\label{eq:lnt}
\end{equation}
notice that $L_{TT}$ and $L_{nT}$ have opposite sign.

On the other hand, the correction $\epsilon$ introduced in equation
(\ref{eq:assume}) is a complicated function of the relativistic parameter
$z$ and $R=\tau^{2}/\tau_{g}^{2}$, which relates the microscopic
and gravitational time scales. In order to gain insight on the magnitude
of the correction $\epsilon$, one can write $\epsilon\left(z,R\right)$.
To do this we insert the values of $L_{TT}$ and $D_{TR}$, given
in equations (\ref{eq:ltt}) and (\ref{eq:dtr}) respectively as functions
of $z$ and the relativistic values of $C_{n}$ and $C_{s}$ in equation
(\ref{eq:assume}) to obtain

\[
\epsilon\left(z,R\right)\equiv R\mathcal{F}\left(z\right),
\]
where 
\begin{equation}
\mathcal{F}\left(z\right)=4z^{3}\mathcal{G}\left(\frac{1}{z}\right)\left(\frac{1}{z}-\frac{1}{\mathcal{G}\left(\frac{1}{z}\right)}\right)\left[\left(\frac{1}{z}-\frac{1}{\mathcal{G}\left(\frac{1}{z}\right)}\right)-\alpha\left(z\right)\right].\label{eq:calF}
\end{equation}

\end{document}